%% file: main.tex
\documentclass[conference]{IEEEtran}
\usepackage{fancyhdr}
\IEEEoverridecommandlockouts
\usepackage{amsmath,amsfonts}

\usepackage{tabularx}
\usepackage{subfig}
\usepackage{graphicx}
\usepackage{physics}
\usepackage{textcomp}
\usepackage[hidelinks]{hyperref}
\usepackage{xcolor}
\usepackage{fancyhdr}
\usepackage{enumitem}
\usepackage{annotate-equations}
\usepackage{afterpage}
\usepackage{array}
\usepackage{multirow}
\usepackage{algorithm}
\usepackage[noend]{algpseudocode}
\usepackage{cite}

\usepackage{bm}
\usepackage{float}
\usepackage[normalem]{ulem}
\usepackage[format=plain,labelfont=bf,labelsep=period,textfont=sl]{caption}
\usepackage{booktabs}
\usepackage{makecell}
\usepackage{tcolorbox,tikz}
\tcbuselibrary{skins}
\tcbuselibrary{breakable}

\newtcolorbox{mybox}[3][]
{
  breakable, 
  enhanced,
  colback         = #2!10,
  colframe        = #2!5,
  boxsep          =-0.5mm,
  borderline west = {1.0mm}{0.05mm}{#3!30}, 
  borderline north = {0.3mm}{0.05mm}{#3!30}, 
  borderline east = {0.3mm}{0.05mm}{#3!30}, 
  borderline south = {0.3mm}{0.05mm}{#3!30}, 
  #1,
}
 
\newcommand{\imetal}{A$_\textsc{Old}$}

\newcommand{\cmetal}{C$_\textsc{Old}$}
\newcommand{\mznmetal}{A$_\textsc{New}$}
\newcommand{\sol}{\textsc{EcoLife}}
\newcommand{\oracle}{\textsc{Oracle}}
\newcommand{\cooopt}{CO$_2$-\textsc{Opt}}
\newcommand{\perfopt}{\textsc{Service}-\textsc{Time}-\textsc{Opt}}
\newcommand{\energyopt}{\textsc{Energy}-\textsc{Opt}}
\newcommand{\new}{\textsc{New-Only}}
\newcommand{\old}{\textsc{Old-Only}}
\newcommand{\econew}{\textsc{Eco}-\textsc{New}}
\newcommand{\ecoold}{\textsc{Eco}-\textsc{Old}}

\makeatletter
\newcommand{\linebreakand}{%
  \end{@IEEEauthorhalign}
  \hfill\mbox{}\par
  \mbox{}\hfill\begin{@IEEEauthorhalign}
}
\makeatother

\usepackage{pifont}
\newcolumntype{P}[1]{>{\centering\arraybackslash}p{#1}}
\colorlet{shadecolor}{gray!20}

\def\BibTeX{{\rm B\kern-.05em{\sc i\kern-.025em b}\kern-.08em
    T\kern-.1667em\lower.7ex\hbox{E}\kern-.125emX}}

\usepackage[utf8]{inputenc}

\begin{document}
\pagenumbering{arabic}
\title{\sol{}: Carbon-Aware Serverless Function Scheduling for Sustainable Computing
\vspace{-3mm}}

\author{\IEEEauthorblockN{Yankai Jiang}
\IEEEauthorblockA{
\text{Northeastern University}\\
\texttt{jiang.yank@notheastern.edu}}
\and
\IEEEauthorblockN{Rohan Basu Roy}
\IEEEauthorblockA{
\text{Northeastern University}\\
\texttt{basuroy.r@northeastern.edu}
}
\linebreakand
\IEEEauthorblockN
{Baolin Li}
\IEEEauthorblockA{
\text{Northeastern University}\\
\texttt{li.baol@northeastern.edu}
}
\and
\IEEEauthorblockN{Devesh Tiwari}
\IEEEauthorblockA{
\text{Northeastern University}\\
\texttt{d.tiwari@northeastern.edu}}
}

\maketitle
\thispagestyle{fancy}
\lhead{}
\rhead{}
\chead{}
\lfoot{\footnotesize{\label{sc24-footer}
This work has been accepted at SC ’24.
}}
\rfoot{}
\cfoot{}
\renewcommand{\headrulewidth}{0pt}
\renewcommand{\footrulewidth}{0pt}
\renewcommand\IEEEkeywordsname{Keywords}
\begin{abstract}
This work introduces \sol{}, the first carbon-aware serverless function scheduler to co-optimize carbon footprint and performance. \sol{} builds on the key insight of intelligently exploiting multi-generation hardware to achieve high performance and lower carbon footprint. \sol{} designs multiple novel extensions to Particle Swarm Optimization (PSO) in the context of serverless execution environment to achieve high performance while effectively reducing the carbon footprint.


\end{abstract}


\input{sections/introduction}

\input{sections/background}
\input{sections/motivation}
\input{sections/design}

\input{sections/methodology}

\input{sections/evaluation}

\input{sections/related_work}
\input{sections/conclusion}
\input{sections/ack}

\bibliographystyle{IEEEtran}
\bibliography{refs}

\end{document}

%% file: sections/introduction.tex
\section{Introduction}
\label{sec:intro}
\noindent\textbf{Motivation and Goals of \sol{}:} Carbon footprint is increasingly becoming one of the most important measures of sustainability of large-scale computing systems. Due to the growing demand for computing in datacenters, the carbon footprint of these large-scale systems is rising~\cite{gupta2021chasing,gupta2022act,eeckhout2022first,bersatti2024quantifying,wu2022sustainable,kaack2022aligning,lin2023adapting,acun2023carbon,li2023toward,licarbon,li2024toward}. Carbon dioxide (CO$_2$) and other greenhouse gases are emitted during manufacturing datacenter hardware (termed as \textit{embodied} carbon footprint), and also during execution of applications on the hardware (termed as \textit{operational} carbon footprint). The embodied carbon footprint is amortized over the lifetime of the hardware, and the operational carbon footprint depends on the energy consumption of the hardware and the \textit{carbon intensity} of the power grid that provides energy to the datacenter. 

\textit{As detailed in Sec.~\ref{sec:motiv}, we observe that datacenter hardware from different generations (old and new hardware) has different proportions of embodied and operational carbon footprints, and combining hardware from different generations has the potential of jointly minimizing both application runtime and carbon footprint.} In fact, this indirectly opens up the opportunity of potentially extending the lifetime of older hardware for higher environmental sustainability of large-scale computing systems. \sol{} leverages this observation and opportunity in designing a scheduling solution for serverless computing. \textit{\sol{} aims to make serverless computing sustainable and high-performant by performing carbon footprint-aware scheduling of serverless functions on hardware from different generations.} 

\vspace{2mm}

\noindent\textbf{Serverless computing and challenges in carbon-aware serverless scheduling:} Serverless computing is gaining wider adoption as a paradigm of cloud computing due to several attractive features like a higher level of resource abstraction from end-users, auto-scaling of resources, and a pay-as-you-go billing model~\cite{jonas2019cloud,hellerstein2018serverless}. Due to these advantages, there is increasing interest in introducing the serverless computing model to the HPC community and workflows~\cite{roy2022mashup,roy2022daydream,basu2023propack,basu2024starship,chard2020funcx}, along with several related efforts in the parallel system's community~\cite{carver2020wukong,skluzacek2019serverless,roy2021characterizing,spillner2018faaster,malawski2020serverless,jiang2017serverless}.

\vspace{1mm}
To make serverless computing high-performant, service providers keep functions alive in the memory of servers so that they are not affected by a start-up overhead, also referred to as the \textit{cold start} of functions. Keeping functions alive consumes resources and energy, which translates to a keep-alive carbon footprint. The summation of the keep-alive carbon footprint and the carbon footprint during execution constitutes the total carbon footprint of a function. Since older hardware often has lower embodied carbon~\cite{Lorenzini_2021}, it can be beneficial to keep functions alive in older hardware but suffer from performance degradation during execution (Sec.~\ref{sec:motiv}). Newer hardware is usually more energy efficient, and hence, results in lower operational carbon --- representing a trade-off between different types of carbon footprint and performance (Sec.~\ref{sec:motiv}). 

Furthermore, different serverless functions need to be \textit{kept alive} for different amounts of time depending on a function's arrival probability. Moreover, the carbon intensity of a power grid varies with time, which has an impact on the operational carbon footprint. \textit{Since the characteristics and invocation patterns of production serverless functions and the carbon intensity vary with time, this makes carbon-aware function scheduling challenging.}

\vspace{1mm}

\noindent\textbf{\sol{}'s Key Contributions:} \sol{} makes the following key contributions. 
\vspace{2mm}

\noindent\textbf{I.} \sol{} is \textit{a novel high-performance and carbon-aware serverless scheduler} that exploits hardware of different generations to improve the sustainability of computing systems (\textit{exploiting the lifetime extension of older-generation hardware}) while achieving high performance. To the best of our knowledge, this is the first work that focuses on reducing the carbon footprint of serverless computing.
\vspace{2mm}

\noindent\textbf{II.} \sol{} introduces novel extensions to the Particle Swarm Optimization (PSO) technique in the context of serverless scheduling. The novel design and implementation of PSO extensions include (a) \textit{perception-response} mechanism to adapt in the dynamic serverless environment, and (b) \textit{function warm pool adjustment} mechanism to intelligently prioritize function keep-alive time and location among multi-generation hardware, in response to varying memory requirements. 
\vspace{1mm}

\noindent\textbf{III.} \sol{} is evaluated using widely-used serverless function invocation trace from Microsoft Azure cloud~\cite{shahrad2020serverless}, and is shown to consistently perform close to the theoretically-optimal (\oracle{}) solution using multiple different generations of hardware and is robust under different scenarios. Our evaluation indicates that \sol{} is consistently within 7.7\% and 5.5\% points from \oracle{} in terms of service time and carbon footprint, respectively. \sol{} is available at \url{https://zenodo.org/records/10976139}.

%% file: sections/background.tex
\section{Background}
\label{sec:bkgd}

\noindent\textbf{Serverless function keep-alive in cloud computing.} The serverless computing model abstracts the cloud computing infrastructure for users, allowing them to upload their code to be executed as stateless functions. In the serverless computing model, cloud providers manage user functions as container images and orchestrate the underlying hardware resources without a need for user intervention. Upon invocation, a function's image is loaded into the server for execution. To enhance efficiency, after execution, the function remains in the server memory for a certain period. The duration during which the function is kept alive in memory is termed the \textit{keep-alive time}, and is controlled by the cloud provider. Keeping functions alive in the memory decreases the chances of a \textit{cold start} of a function, potentially eliminating the need to reload the function into memory. If a function is re-invoked post the keep-alive period, it incurs a \textit{cold start overhead} that requires loading the function in the memory. If a function is re-invoked before the keep-alive period, it receives a \textit{warm start}. The \textit{service time} of a function is comprised of its cold start overhead (zero in the case of a warm start) plus the execution time of the function. Given that the execution times for typical production serverless functions can be comparable to the cold start overhead~\cite{yu2020characterizing,copik2021sebs}, optimizing keep-alive time for serverless functions is an important design consideration.

\vspace{2mm}
\noindent\textbf{Carbon footprint of computing systems.} The carbon footprint encompasses both embodied carbon and operational carbon. Embodied carbon refers to the emissions associated with manufacturing and packaging computer hardware~\cite{lovehagen2023assessing}, such as that from foundries like TSMC. Since this occurs only once, the share of embodied carbon for a traditional (non-serverless) application is proportional to the execution time of the application relative to the lifespan of the device~\cite{gupta2022act,gupta2021chasing}. 

Indeed, as the rapid development in the lithography process continues in hardware manufacturing, advanced hardware has improved capabilities. However, it often comes with larger die sizes, increased core counts, and expanded memory sizes~\cite{Shilov_2019,WikiChip}. Hence, the manufacturing process for such newer-generation hardware often has a higher embodied carbon footprint compared to the older-generation hardware. These carbon emissions generated during manufacturing contribute to the embodied carbon footprint of the hardware and are taken into account throughout its lifespan. 

Unlike embodied carbon footprint, operational carbon footprint refers to the emissions originating from the electricity supplied by grid operators to power the computing infrastructure. It is quantified as the product of the grid’s carbon intensity (gCO$_2$/kWh) and energy usage (kWh). Here, carbon intensity denotes the amount of carbon dioxide emitted per unit of generated energy, and it varies over time. For a traditional (non-serverless) application, the operational carbon footprint includes energy consumed during the execution.

\vspace{2mm}
\noindent\textbf{Carbon footprint estimation for a serverless function.} In contrast to traditional non-serverless functions, the overall carbon footprint of a serverless function is calculated for all three periods: keep-alive period, duration of potential cold-start, and execution time (the first two periods are serverless-specific). The carbon footprint estimation in serverless is composed of \textit{embodied carbon footprint estimation} and \textit{operational carbon footprint estimation}. The embodied and operational carbon footprint of a serverless function accounts for the carbon footprint generated by both the CPUs and DRAMs~\cite{gupta2021chasing,li2023toward,gupta2022act}. Below, we briefly describe how carbon footprint is estimated -- with the acknowledgment that the carbon footprint estimation of serverless functions is non-trivial and has many complex interactions, but the model described below captures the first-order principles and effects. 

First, for the embodied carbon footprint, the attribution of embodied carbon is different during different phases (e.g., keep-alive period and execution time) due to differences in the amount of resources being used. The embodied carbon footprint estimation of DRAM and CPU is accounted for the usage proportion attributed to function $f$. The embodied carbon footprint per unit of time of a DRAM is calculated by dividing the total embodied carbon of the DRAM ($EC_{\text{DRAM}}$) by its lifetime ($LT_{\text{DRAM}}$), and multiplied with the memory usage ratio -- $\frac{M_{f}}{M_{\text{DRAM}}}$. here, $M_f$ is the memory size of the function $f$ and $M_{\text{DRAM}}$ is the size of DRAM. Then, the embodied carbon of DRAM with keep-alive time $k$ and service time $S_{f}$ can be modeled as:

\vspace{-1em}
{\small
\begin{align}
\begin{split}
    \text{DRAM Embodied CO}_{2} = \frac{S_{f}+k}{LT_{\text{DRAM}}}\cdot\frac{M_{f}}{M_{\text{DRAM}}}\cdot EC_{\text{DRAM}}
     \notag
\end{split}
\end{align}}
\label{eq:ecdram}
\vspace{-1em}

During the service period, the entire CPU is assigned to serverless execution. However, during the keep-alive period, one CPU core is preserved to keep the serverless function alive. The embodied carbon per core of a CPU is determined by dividing the total embodied carbon footprint of the CPU ($EC_{\text{CPU}}$) by the number of cores (Core$_\text{num}$). Therefore, the formal expression for the embodied carbon of the CPU can be written as:

\vspace{-1em}
{\small
\begin{align}
\begin{split}
    \text{CPU Embodied CO}_{2} = \frac{S_{f}}{LT_{\text{CPU}}}\cdot EC_{\text{CPU}} + \frac{k}{LT_{\text{CPU}}}\cdot\frac{EC_{\text{CPU}}}{\text{Core}_{\text{num}}} 
     \notag
\end{split}
\end{align}}
\label{eq:eccpu}
\vspace{-1em}

\begin{table}[t]
\centering
\caption{Multi-generation Hardware Pairs Examples.}
\vspace{-2mm}
\scalebox{0.84}{
\begin{tabular}{cccc}
\toprule
\textbf{Pair} & \textbf{Old/New}&\textbf{CPU Model (Year)}& \textbf{DRAM Model (Year)}   \\  
\midrule
\midrule
\multirow{2}{*}{Pair$_A$} & \makecell[c]{A$_\textsc{Old}$} & \makecell[c]{Intel Xeon E5-2686 (2016)}& \makecell[c]{Micron-512 (2018)} \\
\cline{2-4} & \makecell[c]{A$_\textsc{New}$}& \makecell[c]{Intel Xeon Platinum 8252C (2020)} &\makecell[c]{Samsung-192 (2019)}\\ 
\midrule
\multirow{2}{*}{Pair$_B$} & \makecell[c]{B$_\textsc{Old}$}& \makecell[c]{Intel Xeon Platinum 8124M (2017)} &\makecell[c]{Micron-192 (2018)} \\
\cline{2-4} & \makecell[c]{B$_\textsc{New}$}& \makecell[c]{Intel Xeon Platinum 8252C (2020)} &\makecell[c]{Samsung-192 (2019)}\\
\midrule
\multirow{2}{*}{Pair$_C$} & \makecell[c]{C$_\textsc{Old}$}& \makecell[c]{Intel Xeon Platinum 8275L (2019)} &\makecell[c]{Samsung-192 (2019)}\\
\cline{2-4} & \makecell[c]{C$_\textsc{New}$}& \makecell[c]{Intel Xeon Platinum 8252C (2020)} &\makecell[c]{Samsung-192 (2019)}\\
\bottomrule
\end{tabular}}
\vspace{-4mm}
\label{table:server-pairs}
\end{table}

The embodied carbon footprint of hardware is already incurred during manufacturing, but it must be considered and accounted for during the operational period too. Similar to how energy consumption is tracked, the distribution and attribution of the embodied carbon footprint among different applications must be carefully accounted for to inform future planning and potential resource usage.

 Second, the operational carbon footprint includes executing serverless functions during invocation, in addition to the energy required to maintain it in memory during keep-alive. The operational carbon footprint of DRAM can be estimated by multiplying the real-time carbon intensity (CI) with the energy consumption of DRAM ($E_{\text{DRAM}}^{\text{Service}}+E_{\text{DRAM}}^{\text{Keep-alive}}$) during both the service period and keep-alive period. Note that the operational carbon footprint of DRAM incurs a carbon footprint based on the function's share of the overall operational carbon footprint of DRAM. Therefore, the memory usage ratio -- $\frac{M_{f}}{M_{\text{DRAM}}}$ is multiplied to estimate the operational carbon footprint generated by DRAM of a function. The estimation of the CPU's operational carbon footprint is similar to the estimation of embodied carbon. The entire CPU is assigned to serverless function execution, but during the keep-alive period, only one CPU core is used to keep the function alive. These estimations can be formally expressed as:

\vspace{-1em}
{\small
\begin{align}
\begin{split}
\text{DRAM Operational CO}_2 =  \frac{M_{f}}{M_{\text{DRAM}}} \cdot( E_{\text{DRAM}}^{\text{Service}}+  E_{\text{DRAM}}^{\text{Keep-alive}})\cdot \text{CI}\\
\text{CPU Operational CO}_2 = (E_{\text{CPU}}^{\text{Service}} + \frac{E_{\text{CPU}}^{\text{Keep-alive}}}{\text{Core}_{\text{num}} }
)\cdot \text{CI}
     \notag
\end{split}
\end{align}}
\label{eq:operational}
\vspace{-1em}

\label{sec:motiv}
\begin{figure}[t]
    \centering
    \includegraphics[scale=0.51]{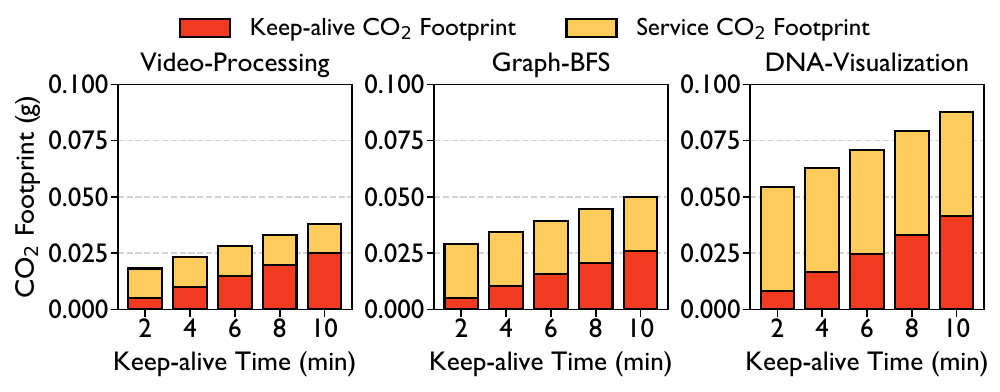}
    \hrule
     \vspace{1mm}
    \caption{The carbon footprint (carbon footprint during keeping-alive and service) for three serverless functions for different keep-alive periods. The contribution of carbon footprint during the keep-alive period toward the overall carbon footprint is significant, esp. as the keep-alive period increases.
    }
    \vspace{-7mm}
    \label{fig:motiv1}
\end{figure}

We acknowledge that a universally accepted methodology is not widely established for the carbon footprint estimation of serverless functions. The approach outlined here is one intuitive method for modeling carbon footprints. Because each period of serverless computing is unique, separately considering each period provides an easy interpretation of the carbon footprint for serverless functions. Incorporating second-level effects (such as storage) can be modeled as an extension by adding the proportional carbon footprint of storage. Our described model does not necessarily favor \sol{} and is only used to demonstrate that carbon savings are possible while achieving high performance. \sol{} primarily focuses on CPU-based systems, which are commonly used in serverless environments~\cite{shahrad2020serverless}. While trends in our motivational observations in Sec.~\ref{sec:motivation} apply to GPUs, too, because of multi-generational trade-offs, we do not directly focus on GPUs. \sol{} can be adapted for multi-generation GPUs using the GPU-specific carbon footprint model and measurement.

Table~\ref{table:server-pairs} shows three old-generation / new-generation pairs to demonstrate that the motivation and key ideas behind \sol{} are not restricted to a single pair, and benefits can be observed over different pairs (Sec.~\ref{sec:eval} confirms this quantitatively). While one cannot practically evaluate all possible multiple generation pairs, entries in Table~\ref{table:server-pairs} were selected to capture three different types of generations (with one, two, and four years of gap representing different lifetimes of the hardware) and where accurate embodied carbon footprint data is available. We anticipate that hardware upgrades can happen in a one to five year timeline, and \sol{} demonstrates how it can be used to leverage the prior generation of hardware to achieve both high performance and a low carbon footprint. 

%% file: sections/motivation.tex
\section{Motivation}
\label{sec:motivation}

\noindent\textbf{Observation.} \textit{Serverless functions generate a significant carbon footprint during their keep-alive period — which is unique compared to the traditional non-serverless computing model, where functions are not kept alive in memory in anticipation of an actual invocation.}

First, we measure the carbon footprint generated during the keep-alive period and the service time of different serverless functions from SeBS benchmark~\cite{copik2021sebs} on \mznmetal{} (Table~\ref{table:server-pairs}). Fig.~\ref{fig:motiv1} shows the trends for three representative functions: video processing, graph search, and DNA visualization. Other functions demonstrate a similar trend, but these functions were selected for motivation as they represent diverse characteristics in terms of computational and memory requirements, and also represent the core of many algorithms. 

From Fig.~\ref{fig:motiv1}, we observe that as the keep-alive period increases, the keep-alive carbon footprint also rises due to the increased embodied and operational carbon emissions. Consequently, its proportion in the total carbon footprint becomes higher. For example, when the keep-alive time is increased from 2 minutes to 10 minutes, the keep-alive carbon footprint of function Graph-BFS has increased from previously constituting 18\% of the total carbon footprint to now 52\% of the total carbon footprint. Fig.~\ref{fig:motiv1} also shows that \textit{the carbon footprint during the keep-alive period can often be higher than the carbon footprint during the actual execution} -- this is because the keep-alive period (typically multiple minutes) is often orders of magnitude longer than the execution time (often millisecond to a few seconds).

\begin{figure}[t]
    \centering
    \includegraphics[scale=0.53]{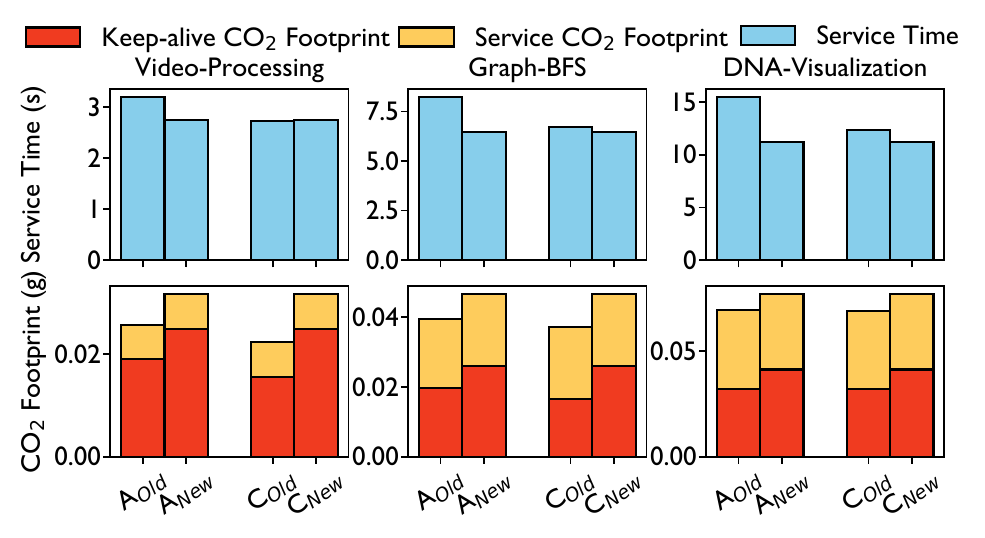}
    \hrule
     \vspace{1mm}
    \caption{The serverless functions can incur a lower overall carbon footprint if kept alive and executed on older-generation hardware due to lower keep-alive carbon footprint (e.g., \imetal{} vs. \mznmetal{}), but they can suffer from performance degradation. However, the impact on performance can be relatively small for some functions with significant savings in carbon footprint (e.g., Graph-BFS on \cmetal{} vs. C$_\textsc{New}$). The keep-alive period is the same and constant (10 minutes) for all cases. }
    \vspace{-0.7cm}
    \label{fig:motiv2}
\end{figure}

\vspace{3mm}
\noindent\textbf{Opportunity.} \textit{The use of relatively older-generation hardware, which inherently has a lower embodied carbon footprint, opens the opportunity to lower the carbon footprint during the keep-alive period. }

We observe this opportunity via comparing the overall service time and carbon footprint across various serverless workloads, using two pairs of older-newer hardware pair as illustrated in Fig.~\ref{fig:motiv2} (Both pair A and C are selected for demonstration). We found that while leveraging relatively older-generation hardware lowers the carbon footprint during the keep-alive period, unfortunately, it results in significant performance (execution time) degradation. For example, as shown in Fig.~\ref{fig:motiv2}, considering executing video processing on \imetal{} and \mznmetal{}, respectively, and keeping the function alive for 10 minutes, using \imetal{} to keep the function alive compared to using \mznmetal{} can save 23.8\% of carbon footprint. However, the execution time of the function increases by 15.9\%. This is because, as expected, older hardware often yields slower performance for many workloads. However, this leads to interesting trade-offs where leveraging older hardware's extended lifetime for carbon footprint benefits competes with the execution time metric.

\begin{figure}[t]
    \centering
    \includegraphics[scale=0.495]{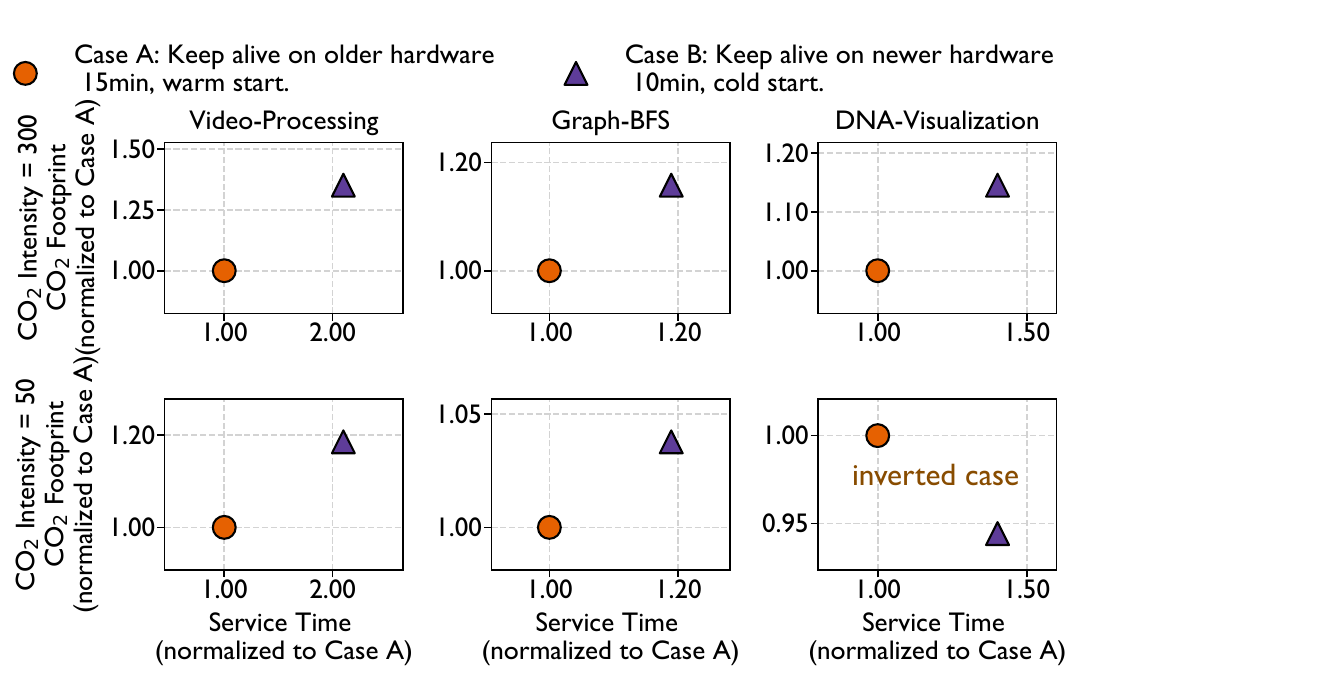}
    \vspace{1mm}
    \hrule
     \vspace{1mm}
    \caption{Trade-off between carbon footprint and service time: A longer keep-alive period on older-generation hardware can potentially reduce both service time and carbon footprint, but the magnitude and feasibility depend on function characteristics and carbon intensity. \textbf {Case A}: keep alive for 15 mins on \cmetal{}, receive a warm start (no cold start overhead) but slower execution. \textbf{Case B}: keep alive for 10 mins on C$_{\textsc{New}}$, and suffer from cold start but faster execution time.}  
    \vspace{-7mm}
    \label{fig:motiv3}
\end{figure}

However, recall that for serverless functions, the metric for performance is service time (not execution time alone). Service time is the sum of the execution time and cold-start overhead (if the function was not warm or kept alive in memory at the time of invocation). \textit{Interestingly, a lower carbon footprint on older hardware during the keep-alive period enables us to afford a longer keep-alive period on older hardware compared to newer hardware under the same or lower carbon footprint budget. This indirectly results in higher chances of warm starts and hence, potentially lower service time even when using older hardware.} However, navigating this trade-off is challenging because the magnitude of the trade-off varies across different functions and hardware generations.

\vspace{3mm}
\noindent\textbf{Challenge.} \textit{To effectively exploit the opportunity identified earlier (trade-off between carbon footprint and service time), one needs to intelligently determine the keep-alive period for different functions on different generations of hardware -- the optimal periods can be different for different functions and vary over time.}

In Fig.~\ref{fig:motiv3}, we perform a comparative experiment to measure the corresponding service time and overall carbon footprint with two testing scenarios (Case A and Case B, described in Fig.~\ref{fig:motiv3}) on two generations of hardware (\cmetal{} vs C$_{\textsc{New}}$) under two carbon intensity (Carbon Intensity = 50, Carbon Intensity = 300). The results provide strong experimental evidence for the previously discussed opportunity. For example, in Fig.~\ref{fig:motiv3} (top row with Carbon Intensity = 300), when the video-processing function is kept alive in memory for a longer time (15 mins) and executed on older hardware (\cmetal{}) with warm start, it leads to a 52.3\% saving in service time and a 14.9\% saving in carbon footprint compared to utilizing shorter keep-alive periods (10 mins) on newer hardware (C$_{\textsc{New}}$) with cold start. This is true for Graph-BFS and DNA visualization functions, too.

We show that utilizing older hardware for extended keep-alive time could potentially reduce carbon emissions while maintaining high performance, because of the increasing possibility of warm starts (hence, eliminating the cold start overhead). Intuitively, when carbon intensity is high, it is worth keeping the function alive on the old-generation hardware (by incurring relatively lower embodied carbon) rather than experiencing the cold start on the new-generation hardware and incurring a high operational carbon footprint during the cold-start period. Essentially, we attempt to eliminate high operational carbon footprint during the cold start on newer hardware by being able to keep the function alive on older hardware -- which incurs lesser embodied carbon and reduces chances of cold start (and hence, better service too). 

However, the magnitude of this benefit can be reduced or absent in some cases when the carbon intensity is very low, and hence, the high operational carbon footprint during cold start on newer generation hardware is less significant compared to embodied carbon on older hardware during longer keep-alive. Our results demonstrate one such case where leveraging older-generation hardware does not always necessarily lead to a lower carbon footprint. The inverted case is shown in Fig.~\ref{fig:motiv3} (bottom) where keeping the DNA-visualization function alive on the older hardware for a longer period improves the service time as before, but may not result in carbon footprint saving -- as alluded earlier, this is because of the impact of temporal variations in the carbon intensity and its impact on the operational carbon footprint. The inversion depends on many factors including carbon intensity, keep-alive period, execution length, cold-start, energy consumption of function, etc., and hence, the inversion point can vary among functions. Energy consumption is different in both scenarios (case A\& B) because case B has longer service time due to cold-start and the DRAM energy may contribute toward the carbon footprint in different amounts for different functions.

In such situations, na\'ively choosing older hardware with a longer keep-alive period does not automatically lead to savings in the carbon footprint and service time and this is why the optimization is challenging. \textit{Choosing keep-alive periods effectively requires carefully considering the carbon intensity of the energy source and adapting to temporal variations of the carbon intensity -- since carbon intensity affects the operational carbon footprint component of the overall carbon footprint (embodied plus operational carbon footprint).}

\begin{figure}[t]
    \centering
    \includegraphics[scale=0.49]{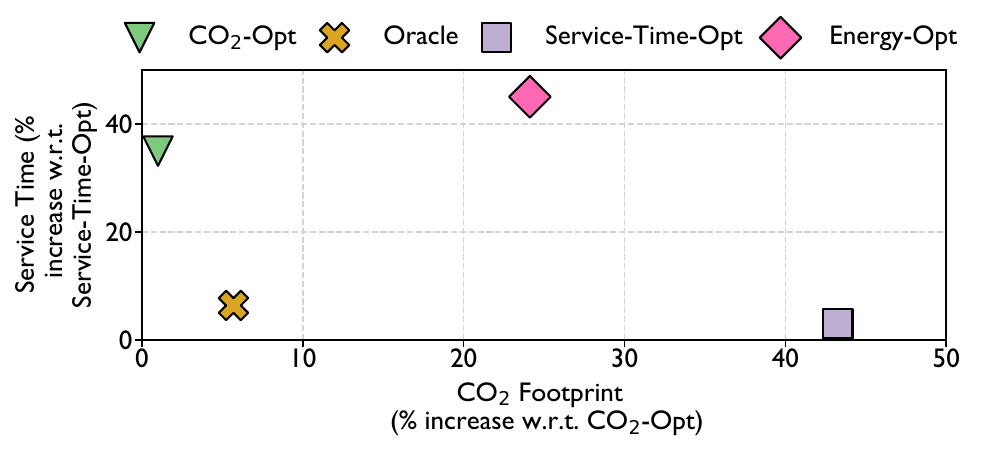}
    \hrule
     \vspace{1mm}
    \caption{\cooopt{}, \perfopt{} and \energyopt{} are far from the \oracle{}. There is an opportunity to jointly optimize the overall carbon footprint and service time, but it is challenging to exploit. (\imetal{} vs \mznmetal{})}
    \vspace{-7mm}
    \label{fig:motiv4}
\end{figure}

\vspace{3mm}
\noindent\textbf{Joint Optimization for Carbon Footprint and Service Time.} Figure~\ref{fig:motiv4} demonstrates the potential for reducing carbon footprint while decreasing service time within the stateless serverless computing environment. The \cooopt{} solution represents the most optimal solution solely focused on minimizing carbon footprint, while the \perfopt{} solution represents the optimal solution for minimizing service time alone. The \oracle{} solution, theoretically optimal, aims to co-optimize both carbon footprint and service time. 

 Note that these three solutions are impractical in real-world systems as they rely on brute-force methods to explore all possible choices, providing only an upper bound for the design reference for \sol{}. The \energyopt{} solution stands as the traditional and naive optimal solution only focused on minimizing energy consumption to indirectly minimize carbon footprint. Notably, although energy consumption primarily contributes to the operational carbon footprint, \energyopt{} solution is far from \cooopt{} solution. This is because energy-aware solutions often overlook the significance of embodied carbon footprint and variations in carbon intensity. 

 Unfortunately, co-optimization of service time and carbon footprint is challenging as shown in Figure~\ref{fig:motiv4} where even the \oracle{} solution is more than 7\% far from the respective \perfopt{} and \cooopt{} solutions. An effective approach for co-optimizing service time and carbon footprint should incorporate the changes in function invocations and carbon intensity. Therefore, it is essential to develop a scheduler capable of adapting to a dynamically changing serverless environment. \sol{} is inspired by these necessities.

%% file: sections/design.tex
\section{Design of \sol{}}
\label{sec:design}
In this section, we first formulate the objective function that \sol{} minimizes. Then, we present the key ideas behind the design of \sol{}.

\subsection{Problem Formulation}
\label{sec:problem}
The goal of \sol{} is to determine the most suitable location (older-generation hardware or newer-generation hardware) and keep-alive periods for serverless functions -- to co-optimize both service time and the carbon footprint. As mentioned in Sec.~\ref{sec:motiv}, the keep-alive period can influence function cold starts, which in turn impacts both service time and carbon footprint. Our optimization can be subdivided into three main components: service time, carbon footprint during function execution, and carbon footprint during the keep-alive period of functions. The following expression shows the general objective function for achieving the optimization goal. 

 \vspace{0.3em}
{\small
\begin{align}
\begin{split}
     \operatorname*{argmin}_{ l\in L, \; k\in \text{KAT}}   \lambda_{s}\eqnmarkbox[darkgray]{p1}{\frac{E[S_{f_{l, k}}]}{S_{f_{\text{max}}}}} + \lambda_{c}\eqnmarkbox[red]{p2}{\frac{E[SC_{f_{l, k}}]}{SC_{f_{\text{max}}}}} + \lambda_{c}\eqnmarkbox[blue]{p3}{\frac{KC_{f_{l,k}}}{KC_{f_{k_{\text{max}}}}}}\notag
\end{split}
\end{align}}
\label{eqal:obj}
\annotate[yshift=0.45em,]{above,left}{p1}{Service Time}
\annotate[yshift=0.45em,xshift = -3em]{above}{p2}{Carbon Footprint During Execution}
\annotate[yshift=-0.05em]{below,left}{p3}{Carbon Footprint During Keep-alive period}
\label{sec:desi_formulation}
\vspace{2mm}

$\lambda_{s}$ and $\lambda_{c}$ are the adjustable parameters to determine the optimization weights on reducing service time and carbon footprint, respectively. $E[S_{f_{l,k}}]$ is the expected value of service time of function $f$, keeping alive on hardware $l$, with $k$ keep-alive time period. $S_{f_{\text{max}}}$ is the maximum service time (the function has a cold start and is executed on the older-generation hardware). Similarly, $E[SC_{f_{l,k}}]$ denotes the expected carbon footprint during the service time of function $f$ when kept alive on hardware $l$ for $k$ keep-alive period. $SC_{f_{\text{max}}}$ represents the maximum carbon footprint during service time. The service time and carbon footprint of the function $f$ account for the service time and carbon generated by the additional latency and delay.

The term $KC_{f_{l,k}}$ is the carbon footprint of the function $f$ during the keep-alive period $k$ on hardware $l$, $KC_{f_{k{\text{max}}}}$ is the maximum carbon footprint during keeping function $f$ alive (function is kept alive on newer-generation hardware).

\sol{} aims to simultaneously determine the keep-alive locations $l$ (older-generation hardware or newer-generation hardware) and the keep-alive periods $k$ (selected from a set of keep-alive period values) for all invoked functions in order to co-optimize all functions, minimizing the overall carbon footprint and service time. To achieve this optimization, the scheduler should have the following design properties: 
(a) Adaptability to variations in function invocation patterns and carbon intensity: \sol{} must be capable of responding to rapidly changing patterns of function invocations and fluctuations in carbon intensity within short time periods. (b) Co-optimization of all invoked functions: Scheduling one serverless function should consider the keep-alive choices of other functions due to limited memory resources. (c) Low decision-making overhead: \sol{} should have low overhead to handle large serverless function invocation loads efficiently.

\subsection{Overview of \sol{}}
\sol{} is the first design using multi-generation hardware and intelligently selected keep-alive period to minimize the carbon footprint while maintaining high performance. \sol{} consists of three key components: function warm pools, the Keeping-alive Decision Maker (KDM), and the Execution Placement Decision Maker (EPDM). These components operate in coordination with one another. \sol{} manages two warm pools that monitor functions that are kept alive in the memory of Docker containers running on hardware spanning two across both old and new generations. Each pool of kept-alive functions has a memory constraint, necessitating \sol{} to ensure that the combined memory usage of all functions kept alive in the warm pool does not exceed the maximum memory capacity available. When the user sends requests of serverless function invocations, \sol{} uses the Keeping-alive Decision Maker to decide the keep-alive time and keep-alive location for every invoked function. If the memory space of hardware is not enough to hold a bursty load of function invocations, \sol{} performs adjustments in the pool of kept-alive functions for better usage of the available memory for incoming new functions that need to be kept alive. Regarding the function execution, \sol{} determines where to execute functions based on the Execution Placement Decision Maker to minimize carbon footprint and service time. Next, we will discuss the detailed design of each of the components of \sol{}, and how they contribute toward meeting the desired design properties discussed previously. 

\subsection{\sol{}'s Keeping-alive Decision Maker (KDM)}
\sol{}'s KDM uses Particle Swarm Optimization (PSO) to determine the keep-alive time of functions. Before going into the basics of PSO and our novel extensions on vanilla PSO to solve \sol{}'s optimization, we discuss the reasons behind using PSO in \sol{}. 

\vspace{1mm}

\noindent\textbf{Why does \sol{} use PSO?} (a) Even the vanilla PSO algorithm is efficient in terms of determining the keep-alive time of serverless functions and has low decision-making overhead~\cite{xu2018reprint}, fulfilling one of the design properties of \sol{}. PSO can rapidly converge to global optima due to its exploration-exploitation balance. Other exploration-exploitation optimization methods, such as reinforcement learning, have a larger overhead and require offline training. This is because PSO relies on simple, pre-defined rules for updating particle positions rather than learning complex policies through trial and error. (b) Due to its strong exploration capabilities, PSO is well suited to perform online optimization. It can continuously adapt to changing conditions and provide near-optimal solutions in dynamic environments, which is needed in a serverless context (one of the desired design properties). In PSO, multiple particles jointly explore the search space, which helps it to converge quickly when variations in system conditions change the optimal solution. Other traditional searching algorithms, such as gradient descent, are slower to adapt to the variations and are usually stuck in the local optima. Deep learning approaches are also not suitable for real-time online optimizations due to high training overhead and training data requirements. (c) In comparison to other heuristic optimization algorithms, such as Artificial Bee Optimization or Grey Wolf Optimization, PSO needs minimal parameter tuning with just three parameters. In our evaluation, we measured that PSO can reduce the carbon footprint by 17.4\% and service time by 7.2\%, compared to the Genetic Algorithm (another closely related nature-inspired optimization technique) with crossover probability of 0.6, mutation probability of 0.01, and population size of 15. Additionally, PSO showed a 6.2\% reduction in carbon footprint and a 13.46\% decrease in service time compared to the Simulated Annealing algorithm, which was set with an initial temperature of 100, a stop temperature of 1, and a temperature reduction factor of 0.9. Next, we briefly discuss the basics of a vanilla PSO optimizer. Thereafter, our extensions on vanilla PSO make \sol{} more suitable in the context of keeping serverless functions alive. 

\vspace{1mm}
\noindent\textbf{Basics of Particle Swarm Optimization.} Particle swarm optimization is a meta-heuristic optimization algorithm inspired by how bird flocks forage. The bird flock effectively locates the best position of food source by sharing information collectively to let other birds know their respective positions. Birds determine whether the position they found is the optimal one and also share information about the best positions of the entire flock. Eventually, the entire bird flock gathers around the best position of food source. PSO utilizes massless particles to simulate birds in a flock, each particle has two attributes: velocity vector and position vector. Velocity represents the speed of movement, and position represents the direction of movement. The quality of each particle's position is determined by the fitness score. At the start of PSO search, $N$ particles will be distributed at random positions in the search space. Particles change their positions in accordance with the following rules after each iteration:

{\small
\begin{align}
\begin{split}
\begin{cases}
    V_{t+1} = \omega*V_{t}+c_{1}r_{1}(X_{pbest}-X_{t})+c_{2}r_{2}(X_{gbest}-X_{t})\\
     X_{t+1} = X_{t} + V_{t+1}
\end{cases}
     \notag
\end{split}
\end{align}}
\label{eq:pso}

$V_{t+1}$ and $X_{t+1}$ represent the updated velocity and position of a particle, respectively, while $V_{t}$ and $X_{t}$ denote the previous velocity and position, respectively. $X_{pbest}$ is the optimal position found by the individual particle, and $P_{gbest}$ is the optimal position found by the entire swarm of particles. $\omega$ serves as the inertia weight, determining how much a particle should adhere to its previous velocity. $c_1$ and $c_2$ are cognitive and social coefficients, respectively, controlling the balance between refining the particle's search results and acknowledging the swarm's search results. $r_1$ and $r_2$ are random numbers uniformly distributed between 0 and 1. These adjustable coefficients regulate the trade-off between exploration and exploitation conducted by the swarm of particles. Next, we discuss the first extension that \sol{} performs on a vanilla PSO that helps it to quickly adjust to changing function invocation characteristics of serverless platforms.

\vspace{1mm}

\begin{figure}[t]
    \centering
    \includegraphics[scale=0.45]{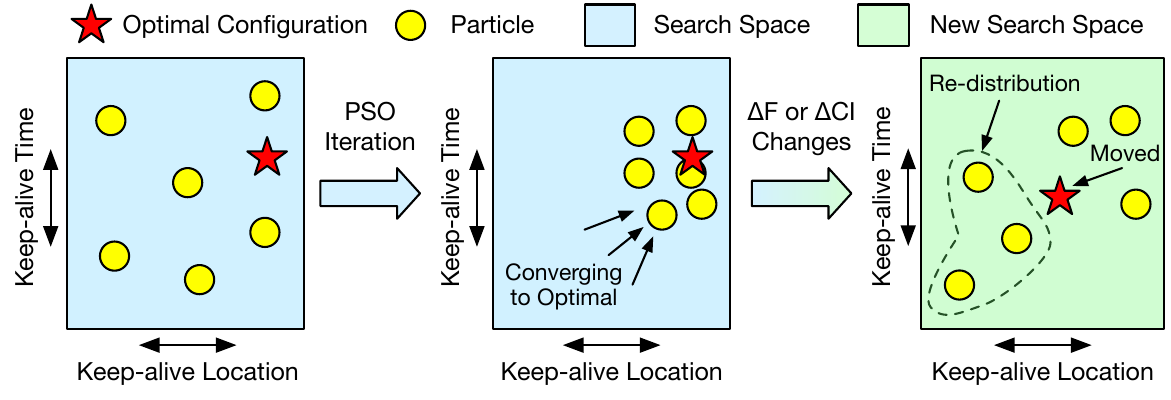}
    \hrule
     \vspace{1mm}
    \caption{Optimization process for \sol{}'s Dynamic PSO (DPSO) -- particles converge to the optimal after movement.}
    \vspace{-7mm}
    \label{fig:design_pso}
\end{figure}

\noindent\textbf{{\sol{}'s Dynamic-PSO.}}
\sol{} constructs a two-dimensional search space for each serverless function to determine the optimal position. One dimension represents two generations of hardware for keeping alive ($l$), while the other dimension covers a set of keep-alive times ($k$). For each new invocation of a serverless function, \sol{} assigns a PSO optimizer and preserves it in order to use it for the next function invocation consisting of parameters for keep-alive location and keep-alive time. However, a vanilla PSO algorithm is not ideally suited for achieving optimal solutions in the serverless environment due to the temporal fluctuations in carbon intensity and function invocations. Note that, configurable weights ($w$, $c_1$, and $c_2$) jointly regulate exploration and exploitation. One intuitive way is to dynamically adjust these weights based on the changes in carbon intensity and function invocation. The weights can be formally expressed as:

\vspace{-1em}
{\small
\begin{align}
\begin{split}
w = w_{\text{max}}\left(\frac{\Delta F}{\Delta F_{\text{max}}} +\frac{\Delta\text{CI}}{\Delta\text{CI}_{\text{max}}} \right)\\
c_1=c_2=c_{\text{max}}\left(1-\frac{\Delta F}{\Delta F_{\text{max}}} -\frac{\Delta\text{CI}}{\Delta\text{CI}_{\text{max}}} \right)
     \notag
\end{split}
\end{align}}
\vspace{-1em}

Here, $\omega_{\text{max}}$ denotes the maximum value of inertia weight, $c_1$ and $c_2$ share the same value, and $c_{\text{max}}$ is the maximum value of the empirical coefficient. $\Delta F$ and $\Delta \text{CI}$ denote the absolute changes of function invocations and carbon intensity, respectively, since the last invocation. $\Delta F_{\text{max}}$ and $\Delta \text{CI}_{\text{max}}$ are the maximum absolute changes in function invocations and carbon intensity across all observation windows so far.

To further enhance PSO's responsiveness to serverless environment variations, \sol{} introduces a \textit{perception-response} mechanism to make PSO adapt to the dynamic environment (visually depicted in Fig.~\ref{fig:design_pso}). This mechanism allows the particle swarm to be dynamically updated in response to environmental changes. If the perception indicates a change in the environment, the particle swarm updates to enlarge the exploration area. Conversely, if there is no perceived change in the environment, updates of the swarm are unnecessary. In \sol{}, perception is represented by changes in $\Delta F$ and $\Delta \text{CI}$. \sol{} detects variations and divides the particle swarm into two halves. One-half is randomly redistributed within the search space, intensifying PSO's exploration and its ability to move past local optima. Meanwhile, the other half of the swarm retains its positions, providing the PSO optimizer with a level of memory, making it easier to find the optimal solution in dynamically changing environments.

\begin{figure}[t]
    \centering
    \includegraphics[scale=0.43]{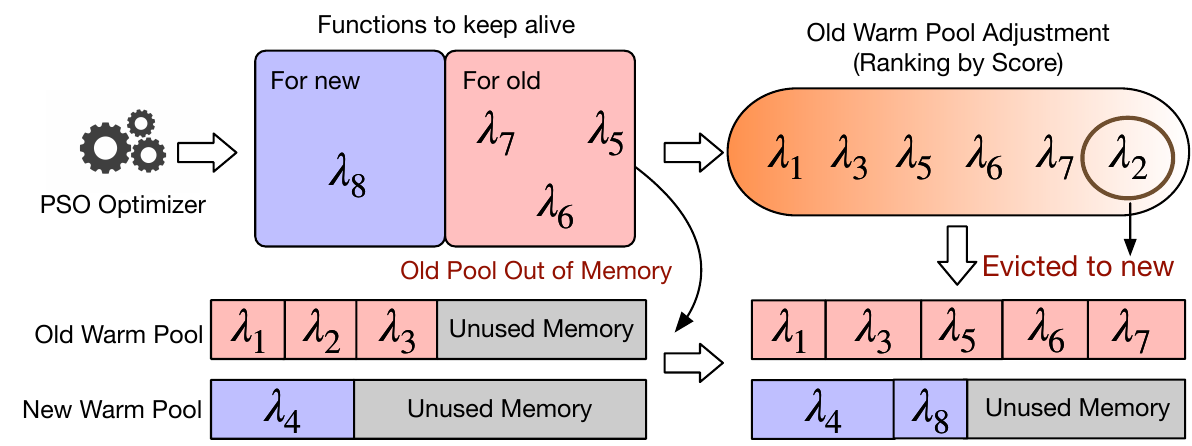}
    \vspace{1mm}
    \hrule
     \vspace{1mm}
    \caption{\sol{} strategically evicts functions from the warm pool when reaching the memory limit. Eviction is based on a priority score, evicted function is kept warm in the other generation's memory if there is enough space. }
    \vspace{-8mm}
    \label{fig:design_pool}
\end{figure}

\vspace{1mm}

\noindent\textbf{\sol{}'s Warm Pool Adjustment.} After collecting all the keep-alive decisions generated by \sol{}'s PSO, functions are designated for keeping alive on either old hardware, or new hardware, otherwise no keep-alive at all. Following these decisions, serverless functions will be kept alive in memory for the entire keep-alive period. However, incoming serverless functions may not be allocated due to memory limitation, despite their greater necessity for being kept alive. This can result in sub-optimal solutions. To address this issue, \sol{} adopts a priority eviction mechanism to sort functions already kept alive in the warm pool as well as those about to be kept alive to find the best arrangement (visually depicted in Fig.~\ref{fig:design_pool}). This is performed by calculating the difference in service time and carbon footprint between cold start and warm start for functions on both old and new hardware to do priority ranking. After performing warm pool adjustment on one type of hardware, functions that are unable to be kept alive due to hardware memory limitation are transferred to another type of hardware for keeping alive, maximizing the utilization of hardware for keeping alive and thus increasing the probability of warm starts. Through warm pool adjustment, \sol{} co-optimizes all serverless functions to reduce service time and carbon footprint, which is the design objective of \sol{}.

\subsection{\sol{}'s Execution Placement Decision Maker}
\sol{} uses the Execution Placement Decision Maker (EPDM) to determine where to execute this function. If the Docker containers on hardware retain this function, it implies that regardless of where the function is executed, it will receive a warm start. EPDM will execute this function on this hardware to avoid the cold start overhead. Else, if this function is not kept alive on only hardware, EPDM's decision will be based on the following scores to determine the optimal execution location:

\vspace{-1em}
{\small
\begin{align}
\begin{split}
f_{\text{score}} = \lambda_{s}\frac{S_{r}}{S_{f_{\text{max}}}}+\lambda_{c}\frac{SC_{r}}{SC_{\text{max}}}
     \notag
\end{split}
\end{align}}
\label{eq:score_exe}
\vspace{-1em}

Here, $r$ denotes the execution location for the function $f$. $S_{f_{\text{max}}}$ and $SC_{\text{max}}$ denotes the maximum service time and carbon footprint.

\begin{algorithm}[t]
\small{
\caption{\sol{}'s Scheduling Framework}
\label{alg:pso}
    \begin{algorithmic}[1]
        \State \textbf{Input:} New invoked functions, initiate search space.
        \State \textbf{Output:} Updated old warm pool $p_{\text{old}}^{\prime}$, old warm pool $p_{\text{new}}^{\prime}$.
        
        \State Get the old warm pool $p_{\text{old}}$ and new warm pool $p_{\text{new}}$.
        \For{every new invoked function $f$}
            \State Execute function $f$ based on the EPDM.
            \State Assign the Dynamic PSO optimizer.
            \For{following invocations of function $f$}
            \State Detect environment variations $\Delta F$ and $\Delta \text{CI}$.
            \State Perform particle re-distribution and movement.
            \EndFor
        \EndFor
        \If{$p_{\text{old}}$ or $p_{\text{new}}$ out of memory}
        \State Perform warm pool adjustment, update $p_{\text{old}}$, $p_{\text{new}}$.
        \State Keep functions alive based on updated warm pools.
        \EndIf
        \State \textbf{else} Keep function alive and update $p_{\text{old}}$, $p_{\text{new}}$.
    \end{algorithmic}
    }
\end{algorithm}

\subsection{\sol{}: Combining All Design Elements}
When a new serverless function is invoked for the first time, \sol{} makes the decision to allocate functions for execution (cold starts), and assign a PSO optimizer for this serverless function. PSO optimizer forms the search space for the function and initializes a number of particles randomly distributed in the space. As a function gets invoked multiple times, \sol{} utilizes the EPDM to determine where to execute this function to avoid cold start overhead based on the warm pools. 

After execution, \sol{}'s PSO detects the changes in the serverless environment (carbon intensity and function invocations), and the optimizer belonging to this function updates the PSO weights ($\omega, c_1, c_2$) according to the changes and randomizes half of the swarm to explore in the search space. 

After performing the particle movement, \sol{} uses the global best position generated by the PSO to decide the keep-alive location and keep-alive period. If there is limited memory for keep-alive, \sol{} performs warm pool adjustment for function keep-alive arrangement. Algorithm~\ref{alg:pso} summarizes the scheduling framework of \sol{}. \sol{} meets all the potential design properties as discussed in Sec.~\ref{sec:problem}.

%% file: sections/methodology.tex
\section{Methodology}
\label{sec:method}

\noindent\textbf{Experimental Setup.} \sol{} evaluation uses two types of testing nodes, \texttt{i3.metal} and \texttt{m5zn.metal}, which are selected from the AWS servers. \texttt{i3.metal} comprises a 2016 released 36 cores Xeon E5-2686 CPU, and a 2018 released 512 GiB Micron DRAM. \texttt{m5zn.metal} equips with a 2020 released 24 cores Xeon Platinum 8252C CPU, and a 2019 released 192 GiB Micron DRAM. This hardware configuration corresponds to Pair A in Table~\ref{table:server-pairs}, used as the default configuration in Sec.~\ref{sec:eval}. Serverless functions are executed in the Docker container, as Docker is widely used in serverless execution~\cite{kuntsevich2018distributed}. Additionally, \sol{} uses an Intel Skylake-SP server with 16 cores, 64 GB memory, and a 4 Gbps network bandwidth as the controller node. We store each serverless function as a Docker image in an S3 bucket. The Docker image will be downloaded from the S3 bucket to the testing node assigned by \sol{} in the control node when execution starts during the simulation campaign.

\vspace{1mm}
\noindent\textbf{\sol{} PSO Setup and Configuration.} We deploy the PSO-based \sol{} on the Intel Skylake-SP server as previously discussed. We assign equal weights to both $\lambda_s$ and $\lambda_c$ ($\lambda_s=\lambda_c=0.5$) to ensure equal optimization of service time and carbon footprint. As for the PSO in \sol{}, $\omega$ ranges from 0.5 to 1, $c_1$ ranges from 0.3 to 1 and $c_2$ ranges from 0.3 to 1. These weights jointly control the exploration and exploitation of PSO, as discussed in Section~\ref{sec:design}. We use 15 particles in the PSO, the number of particles influences the decision-making overhead, and changing the number of particles has negligible influence on the optimization results.

\vspace{1mm}
\noindent\textbf{Evaluated Workloads.} Serverless functions are collected from SeBS benchmark suites~\cite{copik2021sebs}, including various scientific serverless workloads. These functions are invoked following the Microsoft Azure trace~\cite{shahrad2020serverless}. During our trace-driven simulation evaluation, the functions in the Microsoft Azure trace are selected for invocation randomly, but uniformly to ensure representativeness. \sol{} maps all serverless functions to the closest match, considering the memory and execution time.

\vspace{1mm}
\noindent\textbf{Carbon Footprint Estimation and Carbon Intensity.} \sol{} follows the carbon estimation mentioned in Sec.~\ref{sec:bkgd}. \sol{} uses a publicly available dataset~\cite{Davy_2024} and well-established calculation methodologies~\cite{Lorenzini_2021} to determine the total embodied carbon of CPU and DRAM. \sol{} uses a typical four-year lifetime~\cite{shehabi2016united},~\cite{govindan2011benefits} for DRAM and CPU. Carbon intensity within \sol{} is gathered from a widely-used Electricity Maps~\cite{electricitymap}, and expanded to minute intervals to capture the temporal environmental variations. \sol{} primarily utilizes carbon intensity from California Independent System Operator (CISO), where carbon intensity fluctuates by an average of 6.75\% hourly, with a standard deviation of 59.24. Additionally, \sol{} collects carbon intensity from Tennessee (TEN), Texas (TEX), Florida (FLA), and New York (NY) for robustness analysis. The \sol{} utilizes Likwid~\cite{psti} - a simple Linux-based tool suite to read out RAPL~\cite{khan2018rapl} energy information and get info about turbo mode steps on bare-metal machines for energy consumption measurement.

\vspace{1mm}
\noindent\textbf{Relevant and Complementary Techniques.} \sol{} is evaluated to compare with the following schemes:

\noindent\textbf{\new{}, \old{}.} 
\new{}, \old{} follow a ten (10) minutes keep-alive policy of OpenWhisk~\cite{kuntsevich2018distributed}. The \new{} scheme prioritizes the utilization of faster, newer hardware for executing functions under high-performance demands. The \old{} scheme operates in the opposite manner, it always utilizes older-generation hardware for executing functions. It is important to note that utilizing multi-generation hardware to keep functions alive is not a feature introduced in either the \new{} or \old{} scheme.

\noindent\textbf{\cooopt{}, \perfopt{}, and \oracle{}.} \sol{} compares against infeasible solutions, including \cooopt{} (Carbon Footprint Optimal Solution), \perfopt{} (Performance Optimal Solution) and \oracle{} (Best Optimal Solution). These solutions utilize heterogeneous hardware and present the theoretical upper bounds, which are computed via brute-forcing every possible scheduling option for each function invocation. 

\noindent\textbf{\econew{}, \ecoold{}.}
These schemes are static versions of \sol{}, and we use single-generation hardware to schedule functions. \econew{} and \ecoold{} primarily emphasize the determination of keep-alive periods while overlooking the trade-off between older hardware and newer hardware, which is the highlight brought by multi-generation hardware that \sol{} concentrates.

\vspace{1mm}
\noindent\textbf{Figures of Merit.} Carbon footprint and service time are two metrics used to evaluate \sol{}. They are represented as percentages under the \perfopt{} and \cooopt{} (\oracle{} in robustness analysis) to show the increase.

%% file: sections/evaluation.tex
\section{Evaluation}
\label{sec:eval}
In this section, we evaluate the effectiveness of \sol{},
explain its effectiveness, and demonstrate its robustness.
\begin{figure}[t]
    \centering
    \includegraphics[scale=0.48]{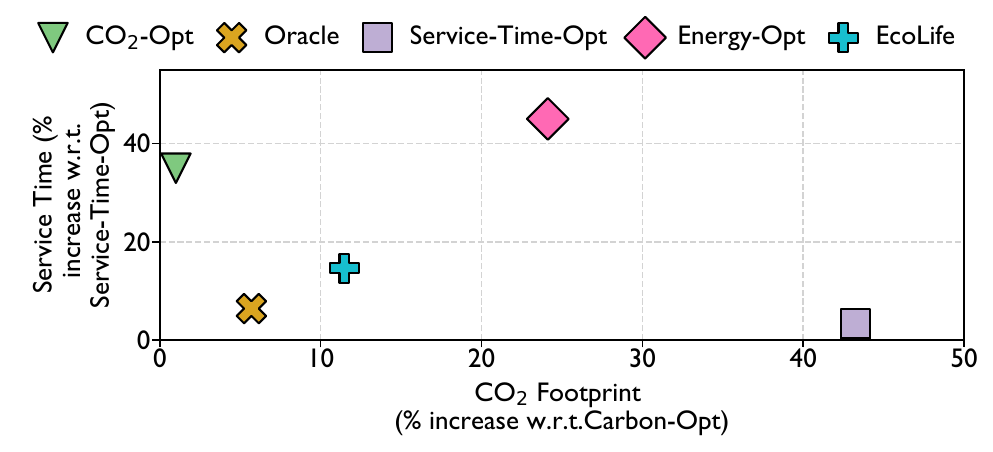}
    \hrule
     \vspace{1mm}
    \caption{\sol{} is closest to the Oracle compared to other relevant techniques.} 
    \vspace{-7mm}
    \label{fig:result}
\end{figure}

\begin{figure}[t]
    \centering
    \includegraphics[scale=0.51]{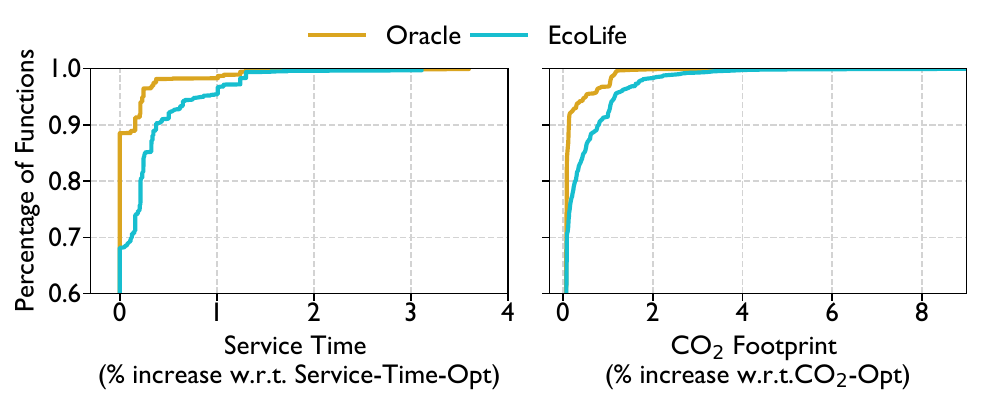}
    \hrule
     \vspace{1mm}
    \caption{Cumulative distribution function for the carbon footprint and service time per function invocation of \sol{} stays close to the \oracle{} for each function invocation.}
    \vspace{-7mm}
    \label{fig:CDF}
\end{figure}

\begin{figure}[t]
    \centering
    \includegraphics[scale=0.53]{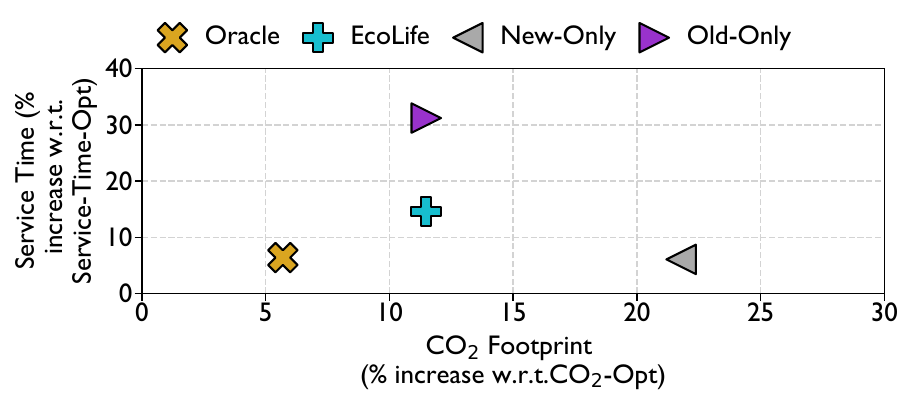}
    \hrule
     \vspace{1mm}
    \caption{
    \sol{} outperforms single-generation only solutions.
    }
    \vspace{-7mm}
    \label{fig:only}
\end{figure}

\subsection{Effectiveness of \sol{}}

    In Fig.~\ref{fig:result}, \sol{} stands out as the closest scheme to the \oracle{} in terms of carbon footprint and service time among all the schemes. Recall from Sec.~\ref{sec:motiv}, \cooopt{}, \perfopt{} and \energyopt{} only minimize carbon footprint, service time and energy consumption respectively, and all of them are significantly far away from the \oracle{}. However, implementing the \oracle{} directly in real-world systems is impractical. \sol{} co-optimize both metrics, while bridging the gap between sustainability and execution performance. Compared to \oracle{}, \sol{} experiences a 7.7\% increase in average service time and a 5.5\% increase in average carbon footprint, respectively. Furthermore, \sol{} is close to \oracle{} from the perspective of individual function invocations. As shown in Fig.~\ref{fig:CDF}, we present the cumulative distribution function (CDF) of service time and carbon footprint respectively, and both service time and carbon footprint remain less than 1\% for each percentile of invoked functions. The service time is the average service time which includes queuing delay, setup delay, cold start (if applicable), and execution time. The P95 latency of \sol{} is within 15\% increase of the service time in \oracle{}. \sol{} decision-making overhead is also low and practical, less than 0.4\% of service time, and 1.2\% of carbon footprint for the invocation loads in the Azure trace. \sol{} achieves scalability by addressing the memory limitation problem of the co-located function with the warm pool adjustment.

As discussed in Sec.~\ref{sec:motiv}, the utilization of multi-generation hardware can bring a high-performance and environmentally friendly serverless execution. In Fig.~\ref{fig:only}, We compare \sol{} with \old{} and \new{} in Sec.~\ref{sec:method} with single-generation hardware under the 10-minute fixed keep-alive policy. While adopting older-generation hardware may reduce carbon emission, \sol{}'s utilization of multi-generation hardware results in a service time saving of 12.7\%. Similarly, although using newer-generation hardware will slightly accelerate the function execution, \sol{} can reduce carbon by 8.6\% with multi-generation hardware. \sol{} is closer to \oracle{} because of its heterogeneity and intelligently selected keep-alive periods. This confirms that \sol{} combines the advantages of both hardware generations to achieve the co-optimization of service time and carbon footprint.

\begin{figure}[t]
    \centering
    \includegraphics[scale=0.54]{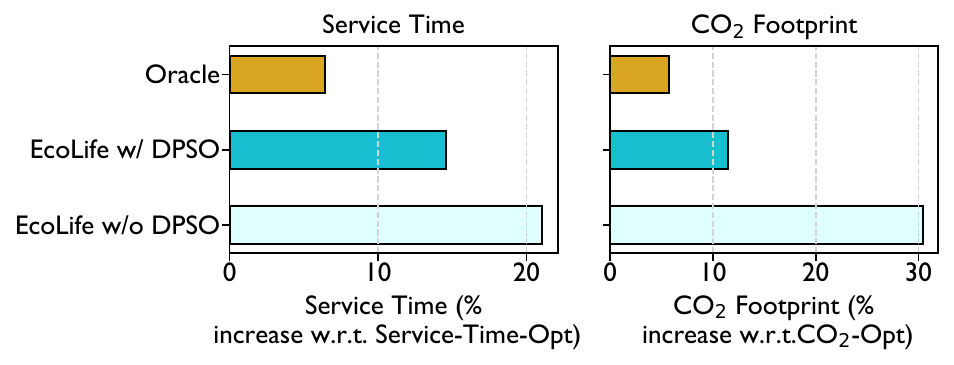}
    \hrule
     \vspace{1mm}
    \caption{Dynamic PSO (DPSO) improves the effectiveness of \sol{}, especially in the terms of carbon footprint.} 
    \vspace{-5mm}
    \label{fig:without_dpso}
\end{figure}

\subsection{Reasons Behind \sol{}'s Effectiveness}
\vspace{-2mm}
\sol{} utilizes a perception-response mechanism to dynamically adapt to the PSO search space based on the changes in function invocations ($\Delta F$) and carbon intensity ($\Delta\text{CI}$), as described in Sec.~\ref{sec:design}. This enables \sol{} to effectively and precisely locate the near-optimal solution. As shown in Fig.~\ref{fig:without_dpso}, \sol{} without dynamic PSO experiences a 5.6\% increase in service time and a 16.9\% increase in carbon footprint. The decisions generated by dynamic PSO impact both the keep-alive period and the keep-alive location, which in turn directly affect the cold start overhead and the carbon footprint. Consequently, without dynamic PSO, the sub-optimal decisions would result in increased service time and carbon footprint.
\begin{figure}[t]
    \centering
    \includegraphics[scale=0.53]{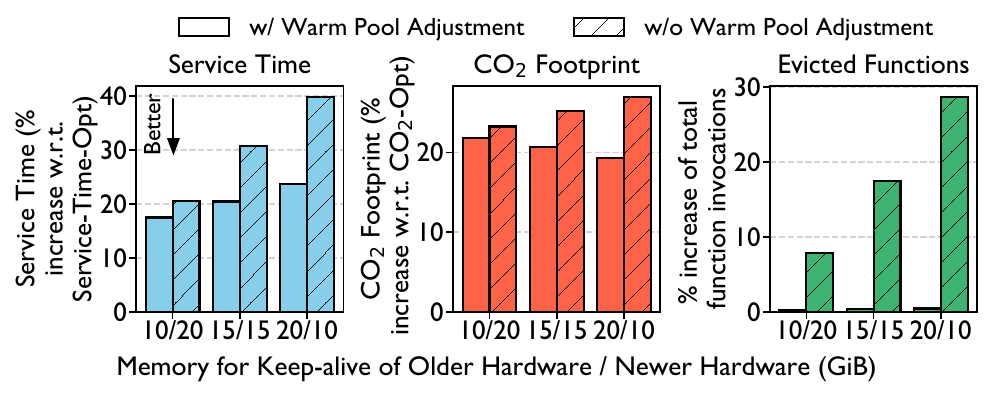}
    \hrule
     \vspace{1mm}
    \caption{\sol{}'s warm pool adjustment strategy is key to its effectiveness.}    %
    \vspace{-0.7cm}
    \label{fig:mem}
\end{figure} 

Warm pool adjustment in \sol{} makes \sol{} effective when memory resources are insufficient to handle numerous serverless functions (Sec.~\ref{sec:design}).

In Fig.~\ref{fig:mem}, we present a comparison of the service time, carbon intensity, and number of evicted functions with and without warm pool adjustment. The old and new hardware keep-alive memory size varies across three combinations, denoted as ``old/new''. The evicted functions are a result of limited memory space in their designated keep-alive hardware. A higher number of evicted functions indicates that the hardware is not utilizing its full potential to keep functions alive, resulting in longer service time because of the more frequent cold starts. As shown in Fig.~\ref{fig:mem}, service time, carbon footprint, and evicted functions with warm pool adjustment are consistently lower than without. For example, with 15GiB memory available on old and new hardware (15/15), warm pool adjustment can save 7.9\% of service time, 3.7\% of carbon footprint, and keep 17\% more functions alive which would otherwise be evicted.

\subsection{Robustness of \sol{}}
\begin{figure}[t]
    \centering
    \includegraphics[scale=0.55]{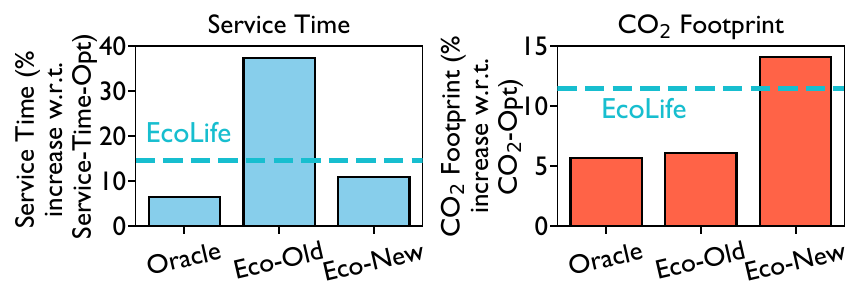}
    \hrule
     \vspace{1mm}
    \caption{\sol{} can be applied to single-generation hardware, but using multi-generation hardware can co-optimize carbon emissions and service time. } 
    \vspace{-0.7cm}
    \label{fig:eco}
\end{figure}

\begin{figure}[t]
    \centering
    \includegraphics[scale=0.53]{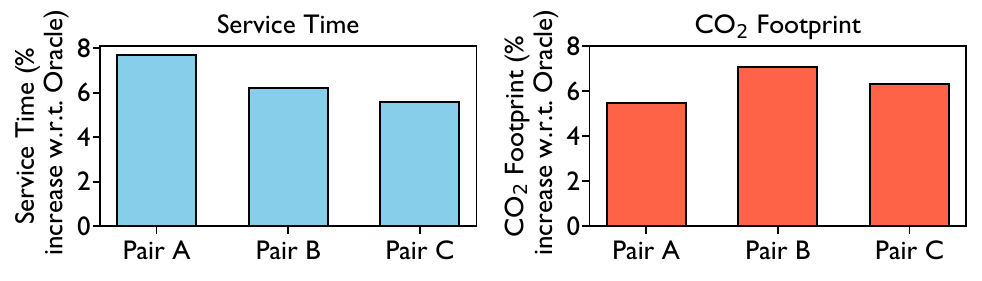}
    \hrule
     \vspace{1mm}
    \caption{\sol{} is effective across different hardware pairs.}
    \vspace{-7mm}
    \label{fig:diff_pairs}
\end{figure}

\sol{} is effective even with single-generation hardware (\ecoold{} or \econew{}), if multi-generation hardware is not available, as demonstrated in Fig.~\ref{fig:eco}. Service time in \ecoold{} and carbon footprint in \econew{} are notably higher compared to the \oracle{}. This is because \oracle{} calculation is based on the multi-generation hardware, which achieves the best balance between minimizing service time and carbon footprint. However, implementing the \oracle{} directly is impractical. \sol{} leverages the advantages of multi-generation utilization and enhances the co-optimization of service time and carbon footprint, offering a viable solution even in the absence of multi-generation hardware. 

\vspace{1mm}
\sol{} is generally applicable to different hardware generation pairs, as we demonstrate its effectiveness against various hardware combinations from Table~\ref{table:server-pairs} in Fig.~\ref{fig:diff_pairs}. Across all hardware generation pairs, \sol{} consistently achieves benefits close to the \oracle{}, as both the service time and carbon footprint remain within a 7.5\% margin to \oracle{}. This demonstrates \sol{}'s ability to flexibly leverage prior-generation hardware to balance service time and carbon footprint, without exploiting specific hardware types.

\vspace{1mm}
\sol{}'s evaluation is focused on demonstrating its effectiveness for a single pair (two generations) to convey the benefit of its key insights. Assuming a five-year lifespan of one generation and alternate-year major hardware upgrades, one would likely expect to predominantly find two or three generations of hardware to be present at a given time in data centers. Three or more generations of hardware also present operational maintainability challenges. Nevertheless, \sol{} can work in the presence of multiple multi-generation pairs, by maintaining multiple warm pools.

\vspace{1mm}
We acknowledge that embodied carbon footprint can have small inaccuracy because the estimation relies on the accuracy of external data sources (e.g. vendor data) and the field is still rapidly evolving with multiple methodological practices. Nevertheless, the benefits of \sol{} remain within 7\% (carbon) and 10\% (service time) of \oracle{} even if we allow a 10\% estimation flexibility range for the embodied carbon footprint. While \sol{} primarily considers the embodied carbon footprint of CPU and DRAM, \sol{} is still effective when considering the embodied carbon footprint of other computer system components, including storage, motherboard, power unit, etc. \sol{} performs within 5.63\% of \oracle{} in carbon footprint and 8.2\% in service time.

\begin{figure}[t!]
    \centering
    \includegraphics[scale=0.50]{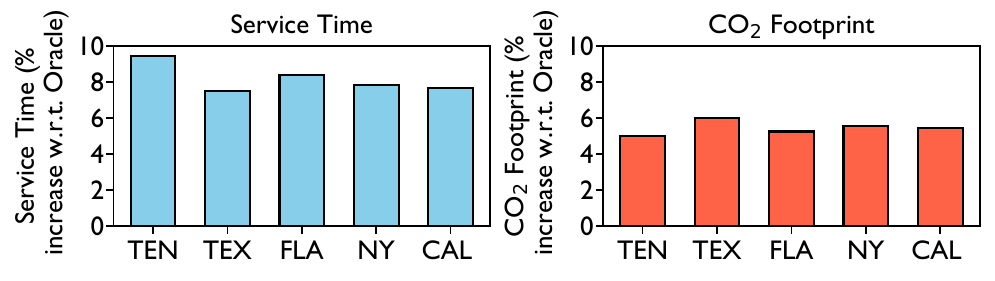}
    \hrule
     \vspace{1mm}
    \caption{\sol{} remains effective and close to \oracle{} across different geographical regions.}    
    \vspace{-7mm}
    \label{fig:diff_ci}
\end{figure}

Finally, we evaluate \sol{}'s effectiveness w.r.t. carbon intensity, as the carbon intensity profile may vary across geographical regions. In Fig.~\ref{fig:diff_ci}, we evaluate \sol{} using carbon intensity data from various regions. The results show that \sol{} remains effective across diverse geographical regions, as it remains within 7\% of the Oracle in terms of service time and 6\% in terms of carbon footprint. This showcases \sol{}'s ability to adapt to various geographical environments and respond to trends in carbon intensity.

\section{Discussion}
\label{sec:discussion}
\vspace{-2mm}

We acknowledge that \sol{}'s unorthodox approach of mixing multi-generational hardware raises several important and interesting considerations. For example, a critical consideration is operational maintainability in a data center environment. We argue that the presence of multi-generational hardware is already natural in today's data center because of portability, compatibility, smooth transition reasons, frequent hardware upgrades, and the need to support diverse customer needs (e.g., Amazon AWS). \sol{} simply exploits that opportunity for saving carbon footprint.

We also highlight that heterogeneity and multi-generation hardware have already been demonstrated to be beneficial from cost and performance perspectives for serverless workloads (e.g., IceBreaker~\cite{roy2022icebreaker}). \sol{} adds one more beneficial dimension -- environmental sustainability to exploit these implicit investments/practices around multi-generation hardware. \sol{} advocates for longer lifetimes for older hardware. Thus, \sol{} opens the avenue for novel research to investigate the trade-off among performance, carbon footprint, lifetime, cost, and maintenance cost.

\sol{}'s idea of lifetime extension for older hardware also has implications for the post-life/disposal carbon footprint of computing hardware. We also recognize that carbon footprint modeling and estimates are currently prone to errors, esp. for embodied carbon. As discussed earlier, \sol{} continues to provide benefits even for a range of estimations, but more efforts are needed to strengthen carbon footprint estimations.

%% file: sections/related_work.tex
\section{Related Work}
\label{sec:related}

\noindent\textbf{Carbon footprint optimizations.} The expansion of cloud and HPC infrastructure has spotlighted the importance of minimizing its carbon footprint, a concern echoed across numerous studies~\cite{gupta2021chasing,gupta2022act,eeckhout2022first,bersatti2024quantifying,li2023toward,wu2022sustainable,kaack2022aligning,lin2023adapting,acun2023carbon}. Efforts to reduce carbon emissions span diverse computation sources, from autonomous vehicles~\cite{sudhakar2022data} and chip design~\cite{chhabria2023towards} to smartphones~\cite{switzer2023junkyard} and the training of large language models~\cite{touvron2023llama,luccioni2023estimating}. Within this context, \sol{} extends this effort into serverless computing, differentiating itself by optimizing the keep-alive strategy of serverless functions for reduced service time and carbon emissions. While cMemento~\cite{kohler2023carbon} introduces carbon-aware memory placement in heterogeneous systems, it does not address the unique challenges of serverless function keep-alive that \sol{} tackles. Although previous research has proposed workload scheduling based on the carbon intensity's temporal and spatial variations~\cite{hanafy2023carbonscaler,li2023clover,liu2011greening,chadha2023greencourier,anderson2023treehouse}, \sol{} innovates by considering serverless functions' execution and keep-alive on multi-generation hardware to promote sustainability.

\vspace{1mm}
\noindent\textbf{Serverless function orchestration.} Serverless computing has emerged as a scalable and efficient service model for cloud users~\cite{castro2019rise,wen2023rise}. Research in this domain has extensively explored optimizations, focusing on cold start mitigation~\cite{oakes2018sock,cadden2020seuss,li2022help,fuerst2021faascache,du2020catalyzer}, hardware resource provisioning~\cite{sahraei2023xfaas,fu2022sfs,jia2021nightcore,yu2022accelerating,bilal2023great,zhou2022aquatope}, stateful execution~\cite{shillaker2020faasm,ding2023automated,roy2022daydream,jia2021boki}, and cost-effectiveness~\cite{zhang2021faster,sadeghian2023unfaasener,eismann2021sizeless}. Amid these developments, a gap remains in addressing the carbon footprint of serverless functions. While Icebreaker~\cite{roy2022icebreaker} and Molecule~\cite{du2022serverless} have explored the use of heterogeneous hardware to enhance serverless function provisioning, they stop short of integrating carbon modeling to harness potential carbon savings. Similarly, energy-efficient solutions for serverless edge computing~\cite{patros2021toward,aslanpour2022energy,byrne2022microfaas} underscore energy savings, which is only one aspect of system carbon footprint. In contrast with all prior works, \sol{} leverages an intelligent keep-alive mechanism on a multi-generation hardware platform, taking the first step towards sustainable serverless cloud computing.

%% file: sections/conclusion.tex
\section{Conclusion}
\label{sec:conclude}

This paper presented \sol{}, a novel placement strategy designed to use multi-generation hardware for optimizing both the carbon footprint and service time of serverless functions. We hope our work will encourage the adaptation of multi-generation hardware in serverless execution environments, promoting more attention to computing sustainability and environmental considerations of large-scale computing systems.

%% file: sections/ack.tex
\newline{\textbf{Acknowledgment.}} We thank the reviewers for their constructive feedback. This work was supported by NSF Awards (2124897, and 1910601) and Northeastern University. This research partially used resources from the Massachusetts Green High Performance Computing Center (MGHPCC). We utilized ChatGPT, an AI language model developed by OpenAI, for partial assistance in drafting the text, and all generated content was thoroughly reviewed and
edited for accuracy.